# Maximum Induced Matching Algorithms via Vertex Ordering Characterizations[*]

Michel Habib[1] and Lalla Mouatadid[2]

1   **IRIF, CNRS & Université Paris Diderot, Paris, France & INRIA Paris, Gang project**
    `Habib@irif.fr`
2   **Department of Computer Science, University of Toronto, Toronto, ON, Canada**
    `Lalla@cs.toronto.edu`

## Abstract

We study the maximum induced matching problem on a graph $G$. Induced matchings correspond to independent sets in $L^2(G)$, the square of the line graph of $G$. The problem is NP-complete on bipartite graphs. In this work, we show that for a number of graph families with forbidden vertex orderings, almost all forbidden patterns on three vertices are preserved when taking the square of the line graph. These orderings can be computed in linear time in the size of the input graph. In particular, given a graph class $\mathcal{G}$ characterized by a vertex ordering, and a graph $G = (V, E) \in \mathcal{G}$ with a corresponding vertex ordering $\sigma$ of $V$, one can produce (in linear time in the size of $G$) an ordering on the vertices of $L^2(G)$, that shows that $L^2(G) \in \mathcal{G}$ - for a number of graph classes $\mathcal{G}$ - without computing the line graph or the square of the line graph of $G$. These results generalize and unify previous ones on showing closure under $L^2(\cdot)$ for various graph families. Furthermore, these orderings on $L^2(G)$ can be exploited algorithmically to compute a maximum induced matching on $G$ faster. We illustrate this latter fact in the second half of the paper where we focus on cocomparability graphs, a large graph class that includes interval, permutation, trapezoid graphs, and co-graphs, and we present the first $\mathcal{O}(mn)$ time algorithm to compute a maximum *weighted* induced matching on cocomparability graphs; an improvement from the best known $\mathcal{O}(n^4)$ time algorithm for the *unweighted* case.

## 1 Introduction

A *matching* in a graph $G(V, E)$ is a subset of edges $M \subseteq E$ where no two edges in $M$ have a common endpoint, i.e. every pair of edges in $M$ is at distance at least one in $G$. An *induced matching* in $G$ is a matching that forms an induced subgraph of $G$, i.e. every pair of edges in the induced matching is at distance at least two in $G$. Induced matching was introduced in [32] by Stockmeyer and Vazirani, as an extension of the matching problem (known as the marriage problem) to the "risk-free" marriage problem. Stockmeyer and Vazirani showed that maximum induced matching is NP-complete on bipartite graphs. The same result was also proven by Cameron in [5]. Since its introduction, the problem has been studied extensively.

---

[*] This work was partially supported by NSERC.

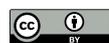
© Michel Habib and Lalla Mouatadid;
licensed under Creative Commons License CC-BY
42nd Conference on Very Important Topics (CVIT 2016).
Editors: John Q. Open and Joan R. Acces; Article No. 23; pp. 23:1–23:17
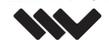
Leibniz International Proceedings in Informatics
LIPICS Schloss Dagstuhl – Leibniz-Zentrum für Informatik, Dagstuhl Publishing, Germany

Induced matchings appear in many real-world applications. For instance, the problem can be used to model uninterrupted communications between broadcasters and receivers [17]. In [1], it was used to model the maximum number of concurrent transmissions in wireless ad hoc networks. In [25], it was used to extract and discover storylines from search results. Induced matchings have also been used to capture a number of network problems, see for instance [20, 2, 14] for network scheduling, gathering, and testing.

The problem is NP-complete even on bipartite graphs of degree three, and planar bipartite graphs [26]. It is also hard to approximate to within a factor of $n^{1-\epsilon}$ and $\Delta_G^{1-\epsilon}$ unless P = NP [13], where $\Delta_G$ is the maximum degree of the graph $G$. In [30], it was shown that the problem is W[1]-hard in general, but planar graphs admit a linear size kernel.

On the tractable side, induced matching is polynomially solvable for a number of graph classes, including trees, weakly chordal, asteroidal-triple free, and circular arc graphs, as well as graphs of bounded clique width [5, 6, 7, 16, 17, 8, 21]. We refer the reader to [13], a survey by Duckworth et al. that contains most of the references and complexity results.

Most of the graph classes for which the problem is tractable have well defined intersection models. One of the main techniques used to show the problem is tractable for a graph class $\mathcal{G}$, is to show that given an intersection representation of a graph $G \in \mathcal{G}$, there exists an intersection representation of a graph $H \in \mathcal{G}$, such that $L^2(G) = H$, where $L^2(G)$ is the square of the line graph of $G$. In other words, one can show that these graph classes are closed under the operation of "taking the square of the line graph" ($L^2(\cdot)$ operation). Since computing a matching (resp. an induced matching) on a graph $G \in \mathcal{G}$ is equivalent to computing an independent set on $L(G)$, the line graph of $G$, (resp. on $L^2(G)$, the square of $L(G)$), by showing closure under $L^2(\cdot)$, the induced matching problem is tractable on $\mathcal{G}$ if and only if computing an independent set is tractable on $\mathcal{G}$.

A *vertex ordering characterization* is an ordering on the vertices of a graph that satisfies certain properties. A graph class $\mathcal{G}$ has a vertex ordering characterization if every $G \in \mathcal{G}$ has a total ordering of its vertices that satisfies said properties. In this work, we use vertex ordering characterizations to show that certain graph classes are closed under $L^2(\cdot)$. In particular, one can observe that lexicographic orderings on the edges of a given vertex ordering of $G$ produces an ordering on the vertices of $L^2(G)$. Since many graph classes are characterized by vertex orderings, and are closed under the square of the line graph operation, it is natural to ask what these orderings on the edges produce as vertex orderings on $L^2(G)$. In [3], Brandstädt and Hoàng showed how to compute perfect elimination orderings of $L^2(G)$ when $G$ is chordal.

In this work we show that almost all forbidden patterns on three vertices are "preserved" under the $L^2(\cdot)$ operation, under two algorithms that compute orderings on $L^2(G)$. This general theorem shows that graph families with certain vertex ordering characterizations are closed under the $L^2(\cdot)$ operation; and these orderings of $L^2(G)$ can be computed in linear time in the size of $G$. This property gives, in our opinion, the most natural way to approach this closure operation, and unfies the results on structural graph classes that have relied on geometric intersection models to show closure. Furthermore, being able to compute vertex orderings directly can be exploited algorithmically, since algorithms on the graph classes covered often rely on their vertex ordering characterizations.

Using two different rules ($\star$ and $\bullet$) to compute these orderings on $L^2(G)$, we show that both the $\star$ and the $\bullet$ rules preserve forbidden patterns in the square of the line graph. As a corollary, we get that threshold, interval, and cocomparability graphs - among other classes - are all closed under $L^2(\cdot)$, and their corresponding vertex ordering characterizations are all preserved under $L^2(\cdot)$. One of the classes we focus on is *cocomparability graphs*, a large graph class that includes interval, permutation, and trapezoid graphs.

In the second half of this work, we present a faster algorithm to compute a maximum *weight* induced matching for cocomparability graphs. Induced matching on cocomparability graphs has been studied first by Golumbic and Lewenstein in [17], then by Cameron in [6], where they both gave different proofs to show that cocomparability graphs are closed under the $L^2(\cdot)$ operation. In [17], they showed that this closure holds for $k$-trapezoid graphs using the intersection representation of $k$-trapezoid graphs; since cocomparability graphs are the union over all $k$-trapezoid graphs, the result holds for cocomparability graphs as well. Whereas in [6], Cameron used the intersection model of cocomparability graphs (the intersection of continuous curves between two parallel lines [18]) to conclude the result directly. Cocomparability graphs are characterized by a vertex ordering known as a *cocomparability* or *umbrella-free* ordering [24]. We use cocomparability orderings and the $L^2(\cdot)$ closure to present a $\mathcal{O}(mn)$ time algorithm to compute a maximum *weighted* induced matching for this graph class, which is an improvement over the $\mathcal{O}(n^4)$ time algorithm for the *unweighted* case - a bound one can achieve by computing $L^2(G)$ and running the algorithm in [11] on it.

The paper is organized as follows: In Section 2, we give the necessary background and definitions. In Section 3, we give the general theorem for a number of graph classes closed under the $L^2(\cdot)$ operation. In Section 4, we present the maximum weight induced matching algorithm and its analysis on cocomparability graphs. We conclude with a discussion on methods that fail, as well as future directions in Section 5.

## 2 Definitions & Preliminaries

We follow standard graph notation in this paper, see for instance [15]. $G = (V, E)$ denotes a *simple graph* (no loops, no multiple edges) on $n = |V|$ vertices and $m = |E|$ edges. $N(v)$ is the *open neighbourhood* of a vertex $v$. The *degree* of a vertex $v$ is $\deg(v) = |N(v)|$. $\Delta_G$ denotes the maximum vertex degree in $G$. We often refer to an edge $(u, v)$ as $uv$. The *distance* between a pair of vertices $u$ and $v$, $\text{dist}_G(u, v)$, is the length of the shortest path between $u$ and $v$ in $G$. The *distance* between a pair of edges $e_1, e_2$, denoted $\text{edist}_G(e_1, e_2)$, is the minimum distance over all shortest paths connecting an endpoint of $e_1$ to an endpoint of $e_2$. The *square* of a graph $G = (V, E)$ is the graph $G^2 = (V, E^2)$ where $uv \in E^2$ if and only if $\text{dist}_G(u, v) \leq 2$. The *chromatic number* of a graph $G$, $\chi(G)$, is the minimum number of colours required to properly colour $G$, i.e, to assign colours to $V$ such that adjacent vertices receive different colours. An *induced subgraph* $H$ of $G$ is a graph $H = (V_H, E_H)$ where $V_H \subseteq V$ and for all $u, v \in V_H, uv \in E$ if and only if $uv \in E_H$. A *matching* $M \subseteq E$ is a subset of edges no two of which share an endpoint. An *induced matching* $M^* \subseteq E$ is a matching in $G$ where every pair of edges in $M^*$ forms an induced $2K_2$, or alternatively every pair of edges in $M^*$ is at distance at least two in $G$. An *independent set* $S \subseteq V$ is a subset of pairwise nonadjacent vertices.

Given a graph $G = (V, E)$, the *line graph* of $G$, denoted $L(G) = (E, L(E))$, is the graph on $m$ vertices, where every vertex in $L(G)$ represents an edge in $G$, and two vertices in $L(G)$ are adjacent if and only if their corresponding edges share an endpoint in $G$. We write $L^2(G) = (E, L^2(E))$ to denote the square of the line graph of $G$.

It is a well known fact that a matching in $G$ is equivalent to an independent set in $L(G)$ [4]. An induced matching on the other hand is equivalent to an independent set in $L^2(G)$ [5]. Two vertices $e_i, e_j$ in $L^2(G)$ are adjacent, i.e. $e_i e_j \in L^2(E)$, if and only if they have one of the configurations in $G$ and $L(G)$ as shown in Fig. 1. In particular, one can see that two vertices are not adjacent in $L^2(G)$ if their corresponding edges induce a $2K_2$ in $G$.

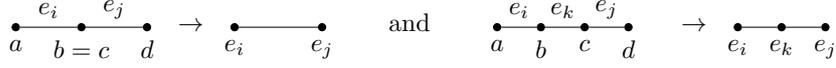

**Figure 1** Configurations of $e_i, e_j \in E$ such that $e_i e_j \in L^2(E)$, and their representation in $L(G)$.

Let $[n] = \{1, 2, \ldots, n\}$. An *ordering* $\sigma$ of $V$ is a bijection $\sigma : V \to [n]$. We write $\sigma = v_1, v_2, \ldots, v_n$. For a pair of vertices $v_i, v_j$, where $i, j \in [n]$ and $i < j$, we write $v_i \prec_\sigma v_j$ or $v_i \prec v_j$ if $\sigma$ is clear in the context.

A *comparability graph* is a graph $G(V, E)$ which admits a transitive orientation of its edges. That is, if two edges $ab, bc \in E$ are oriented $a \to b$ and $b \to c$, then there must exist an edge $ac \in E$ oriented $a \to c$. A *cocomparability graph* is the complement of a comparability graph. Cocomparability graphs are a well studied graph family, see for instance [15]. Given a graph $G = (V, E)$, an ordering $\sigma$ of $G$ is a cocomparability ordering if and only if for every triple $a \prec b \prec c$, if $ac \in E$ then either $ab \in E$ or $bc \in E$, or both. If both $ab, bc \notin E$, we say that the edge $ac$ forms an *umbrella* over vertex $b$. It is easy to see that a cocomparability ordering is just a transitive orientation in the complement. We have the following fact:

▶ **Fact 1.** [24] $G$ is a cocomparability graph iff it admits a cocomparability ordering.

## 3 Vertex Orderings in the Square of the Line Graph

Many well-known classes of graphs can be characterized by vertex orderings avoiding some forbidden patterns, see for example the classification studied in [12] and further studied in [19]. Chordal, interval, split, threshold, proper interval, and cocomparability graphs are a few examples of such graph families. In this section, we show that graphs with certain forbidden induced orderings are closed under the $L^2(\cdot)$ operation. In particular, we show that almost all patterns on three vertices are preserved under $L^2(\cdot)$.

To do so, we construct an ordering on the vertices of $L^2(G)$, and thus on the edges of the original graph $G$, by collecting one edge at a time using different rules; either the $\star$ rule or the $\bullet$ rule. Formally, for a given graph $G = (V, E)$, let $\sigma = v_1, \ldots, v_n$ be a total ordering of $V$. Using $\sigma$, we construct a new ordering $\pi = e_1, \ldots, e_m$ on $E$ as follows: For any two edges $e_i = ab$ and $e_j = uv$ where $a \prec_\sigma b$ and $u \prec_\sigma v$, we place $e_i \prec_\pi e_j$ if:

**Rule ($\bullet$):**   $e_i \prec_\pi e_j \iff a \preceq_\sigma u$ and $b \preceq_\sigma v$

**Rule ($\star$):**   $e_i \prec_\pi e_j \iff \begin{cases} a \prec_\sigma u & \text{if } a \neq u \\ a = u \text{ and } b \prec_\sigma v & \text{o.w.} \end{cases}$

We write $\pi^*(\sigma)$ (resp. $\pi^\bullet(\sigma)$) to denote the ordering constructed using the $\star$ (resp. $\bullet$) rule on $\sigma$. The ordering $\pi^*(\sigma)$ is the lexicographic ordering of $E$ induced by $\sigma$, similar to the one used on chordal graphs in [3]. We will use $\phi^*$ (resp. $\phi^\bullet$) to denote the ordering $\pi^*(\sigma)$ (resp. $\pi^\bullet(\sigma)$) on $L(G)$, *including* the edges $L(E)$; and use $\sigma^*$ (resp. $\sigma^\bullet$) to denote the ordering $\pi^*(\sigma)$ (resp. $\pi^\bullet(\sigma)$) on $L^2(G)$, *including* the edges $L^2(E)$.

▶ **Theorem 2.** *Given a graph $G = (V, E)$, its corresponding $L^2(G) = (E, L^2(E))$, and $\sigma$ an ordering of $V$, if $\sigma$ is $p_i$-free for a pattern $p_i$ in Fig. 2, then $\sigma^\bullet$ is $p_i$-free as well.*

▶ **Theorem 3.** *Given a graph $G = (V, E)$, its corresponding $L^2(G) = (E, L^2(E))$, and $\sigma$ an ordering of $V$, if $\sigma$ is $p_i$-free for a pattern $p_i$ in Fig. 2, then $\sigma^*$ is $p_i$-free as well.*

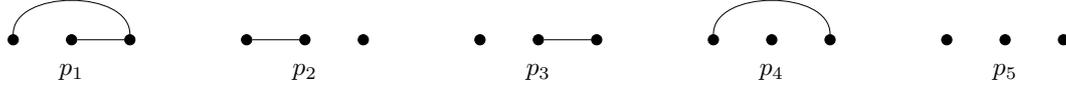

**Figure 2** A list of forbidden patterns on three vertices.

Notice that the pattern $p_4$ forms an umbrella over the middle vertex. Thus the $p_4$-free orderings are precisely cocomparability orderings.

**Proof of Theorem 2.** The proof is by contradiction, where we show if $\sigma^*$ has an induced triple that satisfies a given pattern, then $\sigma$ must also contain such a pattern. Call such a triple $e_1 \prec_{\sigma^*} e_2 \prec_{\sigma^*} e_3$, as shown below:

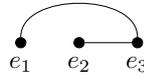

Let $e_1 = ab, e_2 = cd$, and $e_3 = ef$. Without loss of generality, suppose $a \prec_\sigma b, c \prec_\sigma d$, and $e \prec_\sigma f$. Thus $a \preceq_\sigma c \preceq_\sigma e$.

When a triple of vertices $x, y, z$ induces a pattern $p_i$, we write $x, y, z \equiv p_i$. For the ordering $\prec_\sigma$ associated with $\sigma$, we drop the subscript and use $\prec$ instead, whereas we write $\prec_*$ to refer to the ordering $\prec_{\sigma^*}$. Recall that two vertices in $\sigma^*$ are not adjacent iff they induce a $2K_2$ in $G$, and similarly, adjacent vertices in $\sigma^*$ must have $\text{edist}_G \leq 1$ (Fig. 1).

- $p_1$: This pattern produces the following configuration in $\sigma$:

$$ac \wedge ad \wedge bc \wedge bd \notin E \tag{1}$$
$$ae \vee af \vee be \vee bf \in E \tag{2}$$
$$ce \vee cf \vee de \vee df \in E \tag{3}$$
$$a \prec c \preceq e \prec f \tag{4}$$
$$b \prec d \preceq f \tag{5}$$

We begin by showing the all the vertices are distinct.

First, $c \neq e$. Suppose otherwise. Then $ae, be \notin E$ since $ac, bc \notin E$. This implies $af \notin E$ otherwise $a, c = e, f \equiv p_1$. Thus to satisfy (2), $bf \in E$. We place $b$ with respect to $e$, (i.e. $c$). If $e \prec b$ then $e, b, f \equiv p_1$. If $b \prec e$ then $b, e, f \equiv p_1$. Thus $c \neq e$.

Next, we show $d \neq f$. Suppose otherwise. Then $af, bf \notin E$, and $ce \in E$ otherwise $c, e, d = f \equiv p_1$. This in turn implies $ae \notin E$, otherwise $a, c, e \equiv p_1$. Thus to satisfy (2), this leaves $be \in E$. We place $b$ in $\sigma$. If $b \prec c$ then $b, c, e \equiv p_1$. Thus $c \prec b$. If $b \prec e$ then $c \prec b \prec e$ and $c, b, e \equiv p_1$, and if $e \prec b$ then $a, e, b \equiv p_1$. Thus $d \neq f$.

Therefore, all the vertices are distinct and

$$a \prec c \prec e \prec f \tag{6}$$
$$b \prec d \prec f \tag{7}$$

We next show that $ae \notin E$ and $af \notin E$. Suppose first for sake of contradiction that $ae \in E$. Then $ce \notin E$ otherwise $a, c, e \equiv p_1$; and $ce \notin E$ in turn implies $cf \notin E$ otherwise $c, e, f \equiv p_1$. Thus to satisfy (3), $de \vee df \in E$. We place $d$ with respect to $e$. If $e \prec d$ then $de \notin E$ otherwise $c, e, d \equiv p_1$. Thus $df \in E$ but then $e, d, f \equiv p_1$. Thus $d \prec e$. If $de \in E$ then $a, d, e \equiv p_1$. Thus $de \notin E$ but then this leaves $df \in E$ (to satisfy (3)) which forces $d, e, f \equiv p_1$. Thus $ae \notin E$. This implies $af \notin E$ otherwise $a, e, f \equiv p_1$.

Therefore $ae \notin E \wedge af \notin E$, and in order to satisfy (2), we also have $be \in E \vee bf \in E$.

We place $b$ with respect to $e$ first. Either $e \prec b$ or $b \prec e$. If $e \prec b$ then $eb \notin E$ otherwise $a, e, b \equiv p_1$. This forces $bf \in E$ but then $e, b, f \equiv p_1$. Thus $b \prec e$.

We next place $b$ with respect to $c$. Either **(i)** $b \prec c$ or **(ii)** $c \prec b$.

**(i)** Suppose first that $b \prec c$, we thus have $a \prec b \prec c \prec e \prec f$, and $be \vee bf \in E$. Notice that if $bf \in E$ then $be \in E$ otherwise $b, e, f \equiv p_1$. This also implies that $ce, cf \notin E$ otherwise $b, c, e \equiv p_1$ and $b, c, f \equiv p_1$ respectively. Thus to satisfy (3), $de \vee df \in E$. However since $b \prec d \prec f$, it follows that $df \notin E$ otherwise $b, d, f \equiv p_1$. Thus $de \in E$. If $e \prec d$ then $c, e, d \equiv p_1$; and if $d \prec e$ then $b, d, e \equiv p_1$. Thus $bf \notin E$, and it must be that $be \in E$. However this implies $ce \notin E$, otherwise $b, c, e \equiv p_1$. This in turn implies $cf \notin E$ otherwise $c, e, f \equiv p_1$. Thus to satisfy (3), $de \vee df \in E$. We either have $d \prec e$ or $e \prec d$. If $d \prec e$ then $de \notin E$ otherwise $b, d, e \equiv p_1$. This leaves $df \in E$ but then $d, e, f \equiv p_1$. Thus $e \prec d$. This implies $de \notin E$ otherwise $c, e, d \equiv p_1$. Thus $df \in E$ but then $e, d, f \equiv p_1$. Therefore $b \not\prec c$.

**(ii)** $c \prec b$: We have the following ordering $a \prec c \prec b \prec d \prec f$, $b \prec e$, $be \vee bf \in E$, $ae \wedge af \notin E$. We place $e$ with respect to $d$. Either $e \prec d$ or $d \prec e$. Suppose that $e \prec d$. Suppose further that $be \in E$, then $ce \notin E$ otherwise $c, b, e \equiv p_1$, $ce \notin E$ implies $de \notin E$ otherwise $c, e, d \equiv p_1$, and $de \notin E$ implies $df \notin E$ otherwise $e, d, f \equiv p_1$. This leaves $cf \in E$ to satisfy (3), but then $c, e, f \equiv p_1$. Thus $be \notin E$, which forces $bf \in E$ but then $b, e, f \equiv p_1$. Thus $d \prec e$, we have the ordering $a \prec c \prec b \prec d \prec e \prec f$. If $be \in E$ then $de \notin E$ otherwise $b, d, e \equiv p_1$, and $de \notin E$ implies $df \notin E$ otherwise $d, e, f \equiv p_1$. Similarly, $be \in E$ implies $ce \notin E$ otherwise $c, b, e \equiv p_1$, and $ce \notin E$ implies $cf \notin E$ otherwise $c, e, f \equiv p_1$. But this in turn implies $ce \wedge cf \wedge de \wedge df \notin E$, a contradiction to (3). Thus $be \notin E$, which forces $bf \in E$ but then $b, e, f \equiv p_1$. In all cases we get a $p_1$ pattern in $\sigma$.

- $p_2$: This pattern produces the following configurations:

$$ae \wedge af \wedge be \wedge bf \notin E \tag{8}$$
$$ce \wedge cf \wedge de \wedge df \notin E \tag{9}$$
$$ac \vee ad \vee bc \vee bd \in E \tag{10}$$
$$a \preceq c \prec e \prec f \tag{11}$$
$$b \preceq d \prec f \tag{12}$$

Notice that regardless of the total orderings of the vertices, the conditions above are sufficient to conclude that $a, b, f \equiv p_2$.

- $p_3$: This pattern produces the following configurations:

$$ac \wedge ad \wedge bc \wedge bd \notin E \tag{13}$$
$$ae \wedge af \wedge be \wedge bf \notin E \tag{14}$$
$$ce \vee cf \vee de \vee df \in E \tag{15}$$
$$a \prec c \preceq e \prec f \tag{16}$$
$$b \prec d \preceq f \tag{17}$$

Again regardless of the total ordering of the vertices, the conditions are sufficient to conclude that $a, e, f \equiv p_3$.

- $p_4$: This pattern produces the following configurations:

$$ac \wedge ad \wedge bc \wedge bd \notin E \tag{18}$$
$$ce \wedge cf \wedge de \wedge df \notin E \tag{19}$$
$$ae \vee af \vee be \vee bf \in E \tag{20}$$
$$a \prec c \prec e \prec f \tag{21}$$
$$b \prec d \prec f \tag{22}$$

It is sufficient to consider two scenarios, whether $b \prec c$ or $c \prec b$. If $b \prec c$, then to satisfy (20), we have $\alpha \in \{a, b\}, \beta \in \{e, f\}$ such that $\alpha\beta \in E$. However in this ordering we have $a \prec c \prec e \prec f$, thus $\alpha, c, \beta \equiv p_4$. Thus $c \prec b$, but then $a, c, b \equiv p_4$.

- $p_5$: Since $e_1, e_2, e_3$ form a stable set, it follows that $a \prec c \prec e$ and $ac \wedge ae \wedge ce \notin E$, thus $a, c, e \equiv p_5$.

Therefore for all five patters, if there exists a triple $e_1, e_2, e_3 \equiv p_i$ in $\sigma^*$ there must exist a pattern $x, y, z \equiv p_i$ in $\sigma$ as well.

◀

**Proof Of Theorem 3.** We use the same notation as the one used in the proof of Theorem 2 above.

- $p_1$: The configuration in $p_1$ implies the following adjacencies in $G$:

$$ac, ad \notin E \quad \text{and} \quad bc, bd \notin E \tag{23}$$
$$\text{edist}_G(e_1, e_3) \leq 2 \quad \implies \quad ae \vee af \vee be \vee bf \in E \tag{24}$$
$$\text{edist}_G(e_2, e_3) \leq 2 \quad \implies \quad ce \vee cf \vee de \vee df \in E \tag{25}$$

Using the ($\star$) rule, $e_1 \prec_* e_2 \prec_* e_3$ implies $a \prec c \preceq e$. Since $ac \notin E$, it follows that vertices $a$ and $c$ do not share a common neighbour to the right of $c$ in $\sigma$, otherwise the triple would induce a $p_1$ pattern.

We have $c \preceq e$. Suppose first that $c = e$, then $ae, be \notin E$ by (23). By (24), this leaves at least one of $af, bf \in E$. Since $a \prec e \prec f$, it follows that $af \notin E$ otherwise $a, e, f \equiv p_1$. Thus $bf \in E$. Notice that $e \prec b$ for otherwise $b, e, f \equiv p_1$, and also $b \prec f$ otherwise $a, f, b \equiv p_1$. Thus $e \prec b \prec f$, but this implies $e, b, f \equiv p_1$.

Thus $c \neq e$, and $a \prec c \prec e \prec f$.

Next, we show that $ae \notin E$ and $af \notin E$. Suppose first that $ae \in E$, then $ce \notin E$ otherwise $a, c, e \equiv p_1$. Furthermore $ce \notin E$ implies $cf \notin E$ otherwise $c, e, f \equiv p_1$. Thus by (25), either $de \in E$ or $df \in E$ or both. Suppose first $df \in E$. We place $d$ in $\sigma$. Notice that $d \prec f$ otherwise $c, f, d \equiv p_1$. Next we place $d$ with respect to $e$. Either $e \prec d$ or $d \prec e$.

Suppose that $e \prec d$: Then $de \notin E$ otherwise $c, e, d \equiv p_1$. Thus, again by (25), either $cf \in E$ or $df \in E$ or both. If $cf \in E$, then $c, e, f \equiv p_1$ since $c \prec e \prec f$. Thus $cf \notin E$ and $df \in E$. But then $e, d, f \equiv p_1$ since we showed $d \prec f$. In all cases we produce a $p_1$ pattern, therefore $e \not\prec d$.

Thus $d \prec e$; this implies $de \notin E$ otherwise $a, d, e \equiv p_1$. Thus by (25), either $cf \in E$ or $df \in E$ or both. If $cf \in E$ then $c, e, f \equiv p_1$. If $df \in E$ then $d, e, f \equiv p_1$. In all cases we find a $p_1$ pattern in $\sigma$. Therefore $ae \notin E$, and since $e \prec f$, it follows that $af \notin E$ as well, otherwise $a, e, f \equiv p_1$.

We have the following:

$$a \prec c \prec e \prec f \tag{26}$$
$$ae \notin E \quad \text{and} \quad af \notin E \tag{27}$$

We now place vertex $b$ in $\sigma$. Either **(i)** $b \prec c$ or **(ii)** $c \prec b$. Clearly $b \neq e$ since $ab \in E, ae \notin E$.

**(i)** Suppose first $b \prec c$. We have $bc \notin E$, thus $b$ and $c$ do not share any common neighbours to the right of $c$, otherwise $b, c$ and this common right neighbour would induce a $p_1$ pattern. By (24) and (27), either $be \in E$ or $bf \in E$, or both. Notice first that $b \neq f$ by (27). If $be \notin E$ then $bf \in E$ is forced. This implies $b, e, f \equiv p_1$, since by assumption $b \prec c$, and we have $c \prec e \prec f$. Thus $be \in E$. This implies $ce \notin E$ otherwise $b, c, e \equiv p_1$. We place $d$ in the ordering, given that $cd \in E$ and $c \prec d$. Either $d \prec e$ or $e \prec d$. If $d \prec e$ then $b \prec c \prec d \prec e$. And since $bd \notin E$, it follows $de \notin E$ otherwise $b, d, e \equiv p_1$. Thus either $cf \in E$ which implies $c, e, f \equiv p_1$, or $df \in E$ which implies $d, e, f \equiv p_1$. Therefore $e \prec d$. This gives two cases; either $d \prec f$ or $f \prec d$. If the former, then $ed \notin E$ for otherwise $c, e, d \equiv p_1$. Furthermore $ed \notin E$ implies $df \notin E$, otherwise $e, d, f \equiv p_1$. Thus $cf \in E$, but then $c, e, f \equiv p_1$. Thus $f \prec d$, in which case $cf \notin E$ for otherwise $c, e, f \equiv p_1$. Furthermore, $cf \notin E$ implies $fd \notin E$ otherwise $c, f, d \equiv p_1$. Thus $ed \in E$, but then $c, e, d \equiv p_1$. In all cases, we produce a $p_1$ in $\sigma$. Thus $c \prec b$.

**(ii)** We thus have $c \prec b$ and either $e \prec b$ or $b \prec e$. Suppose first that $e \prec b$, then $be \notin E$ otherwise $a, e, b \equiv p_1$. Thus $bf \in E$. If $b \prec f$ then $e, b, f \equiv p_1$, and if $f \prec b$, then $a, f, b \equiv p_1$ since $af \notin E$ by (27). Thus $b \prec e$. We now have the following ordering:

$$a \prec c \prec b \prec e \prec f \tag{28}$$

If $be \notin E$, then $bf \notin E$ otherwise $b, e, f \equiv p_1$. Thus $be \in E$. This implies $ce \notin E$ otherwise $c, b, e \equiv p_1$. This in turn implies $cf \notin E$ otherwise $c, e, f \equiv p_1$. Thus either $de \in E$ or $df \in E$. Notice that in both cases, it cannot be that $f \prec d$ otherwise $ed \in E$ implies $c, e, d \equiv p_1$ and $df \in E$ implies $c, f, d \equiv p_1$. Thus $d \prec f$. If $e \prec d \prec f$, then $ed \notin E$ implies $df \in E$, which in turn implies $e, d, f \equiv p_1$. On the other hand, $ed \notin E$ implies $c, e, d \equiv p_1$. Thus $d \prec e$. If $b \prec d \prec e$, then since $bd \notin E$ and $be \in E$, it follows that $de \notin E$ implies $df \in E$, which in turns leads to $d, e, f, \equiv p_1$. Thus $d \prec b$, in which case $de \notin E$ otherwise $d, b, e \equiv p_1$. $de \notin E$ implies $df \in E$ but then $d, e, f \equiv p_1$. In all cases, we always produce a $p_1$ in $\sigma$.

- $p_2$: Given the ordering in $\sigma^*$ and $\sigma$, it follows that $a \preceq c \prec e \prec f$. We place vertex $b$ in $\sigma$. If $b \prec e$, then $a, b, e \equiv p_2$ since $e_1 e_3 \notin L^2(E)$, and thus $ae, be, af, bf \notin E$. Thus $e \prec b$. We next place vertex $f$ in $\sigma$. If $f \prec b$ then $e, f, b \equiv p_2$ and if $b \prec f$ then $a, b, f \equiv p_2$. Thus the claim of the theorem holds for $p_2$.
- $p_3$: For this pattern, it suffices to notice that $a \prec e \prec f$ always produces a $p_3$ in $\sigma$.
- $p_4$: This configuration in $\sigma^*$ implies the following adjacencies in $G$:

$$ac, ad \notin E \quad \text{and} \quad bc, bd \notin E \tag{29}$$
$$ce, de \notin E \quad \text{and} \quad cf, df \notin E \tag{30}$$
$$\text{edist}_G(e_1, e_3) \leq 1 \implies ae \vee af \vee be \vee bf \in E \tag{31}$$
$$a \prec c \prec e \tag{32}$$

$e_1 e_3 \in L^2(E)$ implies either $e_1$ and $e_3$ are incident edges in $G$ or their distance is at most two in $L(G)$, i.e, $\text{edist}_G(e_1, e_3) \leq 1$. Suppose first that $e_1, e_3$ are incident edges in $G$. This can happen if $e = b$ or $b = f$ since $a \prec c \prec e$.

If $e = b$, we have: $a \prec c \prec e = b \prec f$; and using (29), this implies $a, c, b \equiv p_4$.

If $b = f$, we have: $a \prec c \prec e \prec b = f$, and once again, $a, c, b \equiv p_4$. Thus, $\text{edist}_G(e_1, e_3) \leq 1$. That is, there exists $\alpha \in \{a, b\}, \beta \in \{e, f\}$ such that $(\alpha, \beta) \in E$.

In an attempt to satisfy (31), let's first suppose that $ae \in E$. By (32), $a \prec c \prec e$. By (29, 30), $ae \in E$ would create an umbrella over $c$. Therefore $ae \notin E$. Suppose next that $af \in E$. Since $a \prec c \prec e \prec f$, it follows (using (29, 30)) that $af \in E$ would imply $a, c, f \equiv p_4$. Thus $af \notin E$. Suppose now that $be \in E$. Given that $a \prec b$ and $d \prec e$, we try to place $b$ with respect to $e$. If $e \prec b$ then $a \prec c \prec e \prec b$ and $a, c, b \equiv p_4$ by (29). If $b \prec e$ then either $c \prec b$ or $b \prec c$. If $c \prec b$ then $a, c, b \equiv p_4$. If $b \prec c$ then $b \prec c \prec e$ and by assumption $be \in E$. Thus using (29, 30) $b, c, e \equiv p_4$. In all cases, we produce a $p_4$ if $be \in E$. Therefore $be \notin E$, and to satisfy (31), it remains that $bf \in E$. We place $b$ with respect to $f$. By the same argument above, it must be that $b \prec f$. In fact, $a \prec b \prec c \prec f$ otherwise $a, c, b \equiv p_4$. But $b \prec c \prec f$ and (29, 30) imply $b, c, f \equiv p_4$. Thus $bf \notin E$. We just showed that in all scenarios, condition (31) cannot be satisfied without creating a $p_4$ in $\sigma$. Therefore if $\sigma^*$ has a $p_4$ pattern, then $\sigma$ must have a $p_4$ pattern.

- $p_5$: Since $e_1, e_2, e_3$ form a stable set, it follows that $a \prec c \prec e$ and $ac \wedge ae \wedge ce \notin E$, thus $a, c, e \equiv p_5$.

Once again, for all five patterns, if $\sigma^*$ has a $p_i$ induced pattern then so does $\sigma$.

◀

**Implementation**: Since the $\star$ rule is just a lexicographic ordering on the edges, it is much easier to compute and to store than the $\bullet$ ordering. For this reason, we focus on the $\star$ rule in the remaining of this paper. We begin with the following observation:

▶ **Observation 4.** $\pi^*(\sigma)$ *as computed by the $\star$ rule can be constructed in $\mathcal{O}(m+n)$ time.*

**Proof.** Since the $(\star)$ rule is just a lexicographic ordering on the edges, it suffices to scan the ordering appropriately recording the endpoints of each edge. Formally, suppose $G$ is given as adjacency lists, and let $\sigma = v_1, v_2, \ldots, v_n$ be a total ordering of $G$. For every $w \in V$, we sort the adjacency list of $w$ according to $\sigma$. That is for every pair $v_i, v_j \in N(w)$, if $v_i \prec_\sigma v_j$ then $v_i$ appears before $v_j$ in $N(w)$. This can be done in $\mathcal{O}(m+n)$ time using standard techniques (see for instance [22]). We next construct the ordering $\pi^*(\sigma)$ on the edges of $G$ as follows: Initially $\pi^*(\sigma)$ is empty. We scan $\sigma$ from left to right, for every $v_i$ in $\sigma$, and every neighbour $v_j$ of $v_i$ such that $i < j$, we append $e_k = v_i v_j$ to $\pi^*(\sigma)$. Adding these edges requires scanning $N(w)$ for every $w \in V$. Thus this process takes $\mathcal{O}(m+n)$ time. It is easy to see that this construction satisfies the $(\star)$ rule. We only append $v_i v_j$ for $i < j$ to avoid inserting the same edge twice. The ordering $\pi^*(\sigma)$ we produce at the end of this process is precisely the ordering of the *vertices* of $\pi^*(\sigma), \phi^*$, and $\sigma^*$. Recall that these three orderings differ only in their edge sets and not on the ordering of their vertices. ◀

Therefore if a graph family $\mathcal{G}$ is characterized by the absence of patterns listed in Fig. 2, then if computing an independent set on $G \in \mathcal{G}$ is tractable, and uses the vertex ordering characterization of $\mathcal{G}$, it follows that computing a maximum induced matching is also tractable and reduces to computing an independent set on $L^2(G) \in \mathcal{G}$ using $\sigma^*$.

In this paper, we focus on graph families with forbidden patterns on three vertices (as shown in Fig. 2). To illustrate the consequences of Theorem 3, we list in Table 1 a number of graph families characterized by the absence of the patterns listed in Fig. 2 [4], and Corollary 5 follows immediately. For chordal graphs, Brandstädt and Hoàng gave a stronger result where they showed that not only is $\sigma^*$ a $p_2$-free ordering, but that it is also a lexicographic breadth first search ordering [3].

▶ **Corollary 5.** *Vertex ordering characterizations of threshold, interval, split, cocomparability, and chordal graphs are all closed under the $L^2(\cdot)$ operation, and computing these orderings of $L^2(\cdot)$ can be done in linear time in the size of $G$.*

| $\mathcal{G}$ | Forbidden Patterns |
|---|---|
| Threshold | $p_1$ and $p_2$ |
| Interval | $p_1$ and $p_4$ |
| Split | $p_1$ and $p_3$ |
| Cocomparability | $p_4$ |
| Chordal | $p_1$ |

**Table 1** $G \in \mathcal{G}$ iff $\exists \sigma$ of $G$ that does not have any corresponding induced pattern [12].

## 4 Application: Maximum Weight Induced Matching on Cocomparability Graphs

In this section, we focus on cocomparability graphs. We show how to compute a maximum *weight* induced matching on cocomparability graphs in $\mathcal{O}(mn)$ time, an improvement over $\mathcal{O}(n^4)$ time algorithm for the *unweighted* case. To do so, we use a result we presented in [23], where we give a linear time robust algorithm to compute a maximum weight independent set on cocomparability graphs in linear time. We begin by giving an overview of this algorithm, denoted CCWMIS (Cocomparability Maximum Weighted Independent Set), then present the maximum weight induced matching algorithm and its analysis to achieve the $\mathcal{O}(mn)$ runtime. Thus in the remaining of this section, $G$ is a cocomparability graph and $\sigma$ a cocomparability ordering. By [27], $\sigma$ can be computed in linear time. By Theorem 3 and Observation 4, cocomparability orderings are closed under $L^2(\cdot)$ and can be computed in $\mathcal{O}(m + n)$ time. In particular, notice that the pattern $p_4$ is Fig. 2 is precisely the umbrella forbidden in cocomparability orderings.

### 4.1 Overview of the CCWMIS Algorithm

Let $G = (V, E, w)$ be a *vertex* weighted cocomparability graph, where $w : V \to \mathbb{R}_{>0}$. We compute a cocomparability ordering of $G$, $\sigma = v_1, v_2, \ldots, v_n$. For every vertex $v_i$ in $\sigma$, we assign a set $S_{v_i}$ of vertices. Initially $S_{v_i}$ is empty for all $i \in [n]$. We write $w(S_{v_i})$ to denote the sum of the weights of the vertices in $S_{v_i}$: $w(S_{v_i}) = \sum_{z \in S_{v_i}} w(z)$. We use $\sigma$ to compute a new ordering $\tau = u_1, u_2, \ldots, u_n$ of $G$, by scanning $\sigma$ from *left to right* processing one vertex of $\sigma$ at a time. Initially $\tau_1 = v_1$, and $S_{v_1} = \{v_1\}, w(S_{v_1}) = w(v_1)$. In general, at iteration $i$, when processing a given vertex $v_i$ in $\sigma$, we scan $\tau_{i-1}$ from *right to left* looking for the rightmost nonneighbour of $v_i$ in $\tau_{i-1}$. Let $u$ be such a vertex, if it exists. We construct $S_{v_i} = S_u \cup \{v_i\}$ with $w(S_{v_i}) = w(S_u) + w(v_i)$. If no such $u$ exists, then $S_{v_i} = \{v_i\}$, and $w(S_{v_i}) = w(v_i)$. We show in [23] that the sets $\{S_{v_i}\}_{i=1}^n$ are independent sets.

We proceed to construct $\tau_i$ by inserting $v_i$ into $\tau_{i-1}$. Vertex $v_i$ is inserted into $\tau_{i-1}$ so as to maintain an increasing ordering of the weighted sets $\{S_{v_k}\}_{k=1}^i$. That is, the vertices are ordered in $\tau = u_1, \ldots, u_n$ such that $w(S_{u_i}) \leq w(S_{u_j}), \forall i < j$. When all the vertices of $\sigma$ have been processed, $\tau_n = \tau$ is constructed, we return $S_{u_n}$ as a maximum weight independent set. In [23], we prove the following theorem:

▶ **Theorem 6.** *Let $G = (V, E)$ be a cocomparability graph. Algorithm CCWMIS computes a maximum weight independent set of $G$ in $\mathcal{O}(m + n)$ time.*

---

**Algorithm 1** CCWMIS

---

**Input:** $G = (V, E, w)$ a weighted cocomparability graph where $w : V \to \mathbb{R}_{>0}$
**Output:** A maximum weight independent set together with its weight
 1: Compute $\sigma = v_1, v_2, \ldots, v_n$ a cocomparability ordering of $G$ [27].
 2: **for** $i \to 1$ to $n$ **do**
 3:     $S_{v_i} \leftarrow \{v_i\}$ and $w(S_{v_i}) \leftarrow w(v_i)$
 4: **end for**
 5: $\tau_1 \leftarrow (v_1)$                                                         ▷ Constructing $\tau_i$
 6: **for** $i \to 2$ to $n$ **do**
 7:     Choose $u$ to be the rightmost non-neighbour of $v_i$ with respect to $\tau_{i-1}$
 8:     **if** $u$ exists **then**
 9:         $S_{v_i} \leftarrow \{v_i\} \cup S_u$ and $w(S_{v_i}) \leftarrow w(v_i) + w(S_u)$
10:     **end if**
11:     $\tau_i \leftarrow insert(v_i, \tau_{i-1})$       ▷ Insert $v_i$ into $\tau_{i-1}$ s.t. $\tau_i$ remains ordered w.r.t. $w(S.)$
12: **end for**
13: $z \leftarrow$ the rightmost vertex in $\tau_n$
14: **return** $S_z$ and $w(S_z)$

---

## 4.2 The Weighted Maximum Induced Matching Algorithm (CCWMIM)

Now let $G = (V, E, w)$ be an *edge* weighted cocomparability graph where $w : E \to \mathbb{R}_{>0}$. Thus $L^2(G) = (E, L^2(E), w)$ is a *vertex* weighted cocomparability graph by Theorem 3 and [17, 6]. We compute a maximum weight independent set of $L^2(G)$ as shown in Algorithm 2.

---

**Algorithm 2** Cocomparability Weighted Maximum Induced Matching (CCWMIM)

---

**Input:** $G = (V, E, w)$ an edge weighted cocomparability graph where $w : E \to \mathbb{R}_{>0}$
**Output:** A maximum weight induced matching of $G$
 1: Compute $\sigma = v_1, v_2, \ldots, v_n$ a cocomparability ordering of $G$
 2: Compute $\pi^*(\sigma) = e_1, e_2, \ldots, e_m$ a cocomparability ordering of $L^2(G)$ using the $(\star)$ rule.
    ▷ The ordering only, not the square edges
 3: Use Algorithm 1 and $\pi^*(\sigma)$ to compute a maximum weight independent set of $L^2(G)$

---

By Theorem 6, Algorithm CCWMIS takes $\mathcal{O}(m + n)$ time. Thus, CCWMIS will take $\mathcal{O}(|E| + |L^2(E)|)$ time on $L^2(G)$. When $G$ is dense, CCWMIS on $L^2(G)$ takes $\mathcal{O}(n^4)$ time.

Before giving a careful implementation and analysis to achieve $\mathcal{O}(mn)$ running time, we illustrate Algorithm 2 in Fig. 3, which shows an edge weighted cocomparability graph $G = (V, E, w)$, $\sigma$ a cocomparability ordering of $G$, $\phi^*$ an ordering of $L(G)$ constructed by the $\star$ rule, and $\sigma^*$ the corresponding ordering on the vertices of $L^2(G)$. Table 4 shows the step by step construction of $\tau$, using $\pi^*(\sigma)$ as the ordering computed in Step 1 of Algorithm CCWMIS.

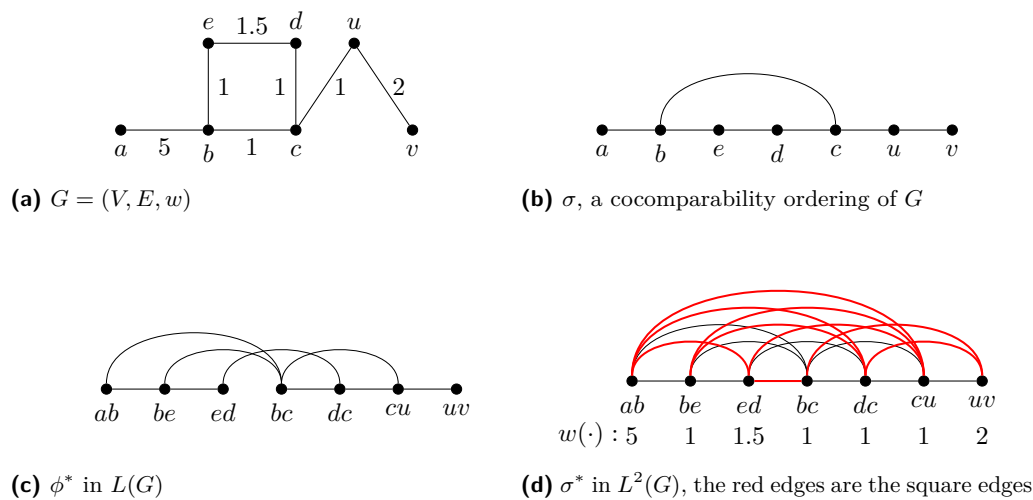

**(a)** $G = (V, E, w)$

**(b)** $\sigma$, a cocomparability ordering of $G$

**(c)** $\phi^*$ in $L(G)$

**(d)** $\sigma^*$ in $L^2(G)$, the red edges are the square edges

**Figure 3** (a) is an edge weighted cocomparability graph, (b) is $\sigma$ a cocomparability ordering of $G$, (c) is a vertex ordering of $L(G)$ produced by the ($\star$) rule, and (d) the cocomparability ordering of $L^2(G)$, produced by the ($\star$) rule.

| $e_i$ | $u$ | $S_{e_i}$ | $w(S_{e_i})$ | $\tau_i$ |
|---|---|---|---|---|
| $e_1 = ab$ | - | $\{e_1\}$ | 5 | $e_1$ |
| $e_2 = be$ | - | $\{e_2\}$ | 1 | $e_2, e_1$ |
| $e_3 = ed$ | - | $\{e_3\}$ | 1.5 | $e_2, e_3, e_1$ |
| $e_4 = bc$ | - | $\{e_4\}$ | 1 | $e_2, e_4, e_3, e_1$ |
| $e_5 = dc$ | - | $\{e_5\}$ | 1 | $e_2, e_4, e_5, e_3, e_1$ |
| $e_6 = cu$ | - | $\{e_6\}$ | 1 | $e_2, e_4, e_5, e_6, e_3, e_1$ |
| $e_7 = uv$ | $e_1$ | $\{e_1, e_7\}$ | 7 | $e_2, e_4, e_5, e_6, e_3, e_1, e_7$ |

**Figure 4** A step by step illustration of Algorithm 1 on the graph in Figure 3. The algorithm returns $S_z = S_{e_7} = \{e_7, e_1\}$ of maximum weight 7.

## 4.3 Implementation & Analysis of CCWMIM

Suppose the graph $G = (V, E, w)$, where $w : E \to \mathbb{R}_{>0}$, is given as adjacency lists. We compute $\sigma = v_1, \ldots, v_n$ in $\mathcal{O}(m + n)$ time using the algorithm in [27]. We construct $\pi^*(\sigma)$ in $\mathcal{O}(m + n)$ time using Observation 4.

Notice that we cannot use $\phi$ as input for the CCWMIS algorithm, since $\phi$ is not necessarily a cocomparability ordering. In fact, $L(G)$ is not necessarily a cocomparability graph; just consider the line graph of any large clique $K_{p>4}$. Notice also that the square edges in $\sigma^*$ are necessary for Step 7 of the algorithm, when looking for a rightmost nonneighbour in $\tau_{i-1}$.

We begin by looking at forbidden configurations of induced $2K_2$s in cocomparability orderings. Let $\sigma = v_1, \ldots, v_n$ be a cocomparability ordering. Let $e_i = ab$ and $e_j = uv$ be two edges that induce a $2K_2$ in $G$. Without loss of generality, suppose $a \prec_\sigma b$ and $u \prec_\sigma v$. Since $\sigma$ is a cocomparability ordering, the configurations of $e_i, e_j$ that have either $a \prec u \prec b \prec v$ or $a \prec u \prec v \prec b$ as orderings cannot occur in $\sigma$, for otherwise $\sigma$ would have an umbrella. This leaves the following configurations of the edges without umbrellas: $a \prec b \prec u \prec v$ or $u \prec v \prec a \prec b$.

Without loss of generality, suppose $a \prec_\sigma b \prec_\sigma u \prec_\sigma v$. Using the $(\star)$ rule, this configuration always forces $e_i \prec_\pi e_j$, i.e. $ab \prec_\pi uv$. Therefore, when we run Algorithm CCWMIS on $\pi^* = e_1, \ldots, e_m$, we process elements of $\pi^*$ from right to left, and thus we process $e_i = ab$ before processing $e_j = uv$.

Let $\tau = f_1, \ldots, f_m$ be the new ordering being constructed by the algorithm CCWMIS using $\pi^*$ as the ordering computed in Step 1. Initially, as per the algorithm, $\tau_1 = e_1$. In general, at iteration $i$, let $\tau_{i-1} = f_1, \ldots, f_{i-1}$ be the ordering constructed thus far. Suppose $e_i$ is the edge being processed. In Step 7 of Algorithm 1, looking for the rightmost nonneighbour of $e_i$ in $\tau_{i-1}$ is equivalent to looking for an edge $e$ that forms an induced $2K_2$ with $e_i$ in $\sigma$, such that $e$ is to the left of $e_i$ in $\sigma$. When processing vertex $e_i$ in $\pi^*$, we scan $\tau_{i-1}$ to find the rightmost nonneighbour of $e_i$ in $\tau_{i-1}$. Suppose such a vertex exists, and call it $f_j$. Since we are working in $L^2(G)$, to check if two vertices in $L^2(G)$ are adjacent, we need to check whether these edges are incident in $G$, or are at distance at most two in $L(G)$, as shown in Fig. 1. We proceed as follows.

Both $\sigma$ and $\pi^*$ are implemented using doubly linked lists. We construct three arrays $A$, $B$ and $F$ of sizes $n, n, m$ respectively. All three arrays are initialized to zero; $A[t] = B[t] = 0, \forall t \in [n]$ and $F[i] = 0, \forall i \in [m]$.

Every vertex $v_t$ in $\sigma$ has a pointer to $A[t]$ and $B[t]$. Similarly, every vertex $e_i$ in $\pi^*$ has a pointer to $F[i]$. We sometimes abuse notation and talk about $A[w]$ to mean the position in array $A$ that vertex $w$ in $\sigma$ points to. Furthermore, when we talk about vertex $e_i = uv$ in $\pi^*$, we always assume that $u \prec_\sigma v$.

For every vertex $e_i = v_t v_k$ in $\pi$, its corresponding entry $F[i]$ has four pointers $p_i^t, p_i^k, q_i^t, q_i^k$ that point respectively to $A[t], A[k]$ and $B[t], B[k]$. When processing vertex $e_i$, where $e_i = ab$, we update $A$ as follows: For every neighbour $z$ of vertex $a$, we set $A[z] = i$. Similarly, for every neighbour $z$ of vertex $b$, we set $B[z] = i$. These updates to arrays $A$ and $B$ guarantee that every nonneighbour $w$ of $a$ has $A[w] \neq i$ and every nonneighbour $w$ of $b$ has $B[w] \neq i$. Therefore, for every edge $v_t v_k$ in $G$ that forms an induced $2K_2$ with $ab$, the following (†) condition holds: $A[t] \neq i \wedge A[k] \neq i \wedge B[t] \neq i \wedge B[k] \neq i$ (†).

Thus, in order to find the rightmost nonneighbour of $e_i$ in $\tau_{i-1}$, we scan $\tau_{i-1}$ from right to left, and for every vertex we encounter $f_j = v_t v_k$, we check if one of $A[t], A[k], B[t], B[k]$ is equal to $i$. We return the first vertex in $\tau_{i-1}$ we encounter whose endpoints in $G$ satisfy condition (†) above as the rightmost nonneighbour of $e_i$ in $\tau_{i-1}$. Updating arrays $A$ and $B$ requires $\mathcal{O}(\deg(a) + \deg(b))$ time. When scanning $\tau_{i-1}$, for every vertex $f_j = v_t v_k$ in $\tau_{i-1}$, we use the pointers $p_j^t, p_j^k, q_j^t, q_j^k$ in $F[j]$ to access $A[t], A[k], B[t], B[k]$. Checking these four entries takes constant time using the pointers provided.

It remains to analyze the number of constant checks we do, i.e. how many $f_j$ vertices we check. In particular, this reduces to bounding the degree of $e_i$ in $L^2(G)$.

Let $\deg_1(e_i)$ denote the degree of $e_i$ in $L(G)$, and $\deg_2(e_i)$ denote the degree of $e_i$ in $L^2(G)$. We have the following:

▶ **Claim 7.** for a given edge $e_i = ab$, we have

$$\deg_2(e_i) \leq \sum_{\substack{v:av \in E \\ v \neq b}} \deg(v) + \sum_{\substack{v:bv \in E \\ v \neq a}} \deg(v)$$

**Proof.** It is clear that for a given edge $e_i = ab$, $\deg_1(e_i) = \deg(a) + \deg(b) - 2$. On the other hand, when computing $\deg_2(e_i)$, we take into account the degree of any vertex at distance at most two from either $a$, or $b$ in $G$. In particular, the following holds:

$$\deg_2(e_i) \leq \deg_1(e_i) + \sum_{\substack{v:av\in E \\ v\neq b}} (\deg(v) - 1) + \sum_{\substack{v:bv\in E \\ v\neq a}} (\deg(v) - 1)$$

$$\leq \deg(a) + \deg(b) - 2 + \left[\sum_{\substack{v:av\in E \\ v\neq b}} \deg(v)\right] - \deg(a) + 1 + \left[\sum_{\substack{v:bv\in E \\ v\neq a}} \deg(v)\right] - \deg(b) + 1$$

$$\leq \sum_{\substack{v:av\in E \\ v\neq b}} \deg(v) + \sum_{\substack{v:bv\in E \\ v\neq a}} \deg(v)$$

The first inequality avoids counting edges twice, in particular if $a, b$, and $v$ form a triangle. The -1s in the first equality is to avoid counting the edge $av$ in $\deg(v)$, for every $v \in N(a)$, similarly for $b$. The +1s in the second equality is for not counting edge $ab$ for both $a$ and $b$ in $\deg(a)$ and $\deg(b)$. ◀

When scanning $\tau_{i-1}$ to find the rightmost nonneighbour of $e_i$, we check $\mathcal{O}(\deg_2(e_i))$ vertices, each check takes constant time using arrays $A, B$, and $F$. Since the weights are positive, $w(S(e_i)) = w(S(f_j)) + w(e_i) > w(S(f_j))$ if such an $f_j$ exists, and thus $f_j \prec_\tau e_i$. Therefore, inserting $e_i$ into $\tau_{i-1}$ to create $\tau_i$ will also take $\mathcal{O}(\deg_2(e_i))$ time.

Summing over all vertices in $\pi$, of which there are $m = |E|$, we have $\mathcal{O}(m + \sum_{e_i} \deg_2(e_i))$. It remains to bound $\sum_{e_i} \deg_2(e_i)$.

▶ **Claim 8.** $\sum_{e_i} \deg_2(e_i) \leq \mathcal{O}(mn)$.

**Proof.** By Claim 7, we have:

$$\sum_{e_i} \deg_2(e_i) \leq \sum_{e_i=(a,b)} \left[\sum_{\substack{v:av\in E \\ v\neq b}} \deg(v) + \sum_{\substack{v:bv\in E \\ v\neq a}} \deg(v)\right]$$

For a given vertex $v$, $\deg(v)$ is used $\deg(v)$ time, one for every edge incident to $v$, thus

$$\sum_{e_i} \deg_2(e_i) = \sum_{e_i=(a,b)} \left[\sum_{\substack{v:av\in E \\ v\neq b}} \deg(v) + \sum_{\substack{v:bv\in E \\ v\neq a}} \deg(v)\right]$$
$$\leq \deg(v_1) \cdot \deg(v_1) + \ldots + \deg(v_n) \cdot \deg(v_n)$$
$$\leq \deg(v_1) \cdot \Delta_G + \ldots + \deg(v_n) \cdot \Delta_G \leq 2m \cdot \Delta_G \leq \mathcal{O}(mn)$$

◀

Therefore, the total running time is $\mathcal{O}(m + mn) = \mathcal{O}(mn)$. The correctness and robustness of the algorithm follows from Theorem 3 as well as the correctness and robustness of Algorithm 1, which we give in [23]. We conclude with the following theorem:

▶ **Theorem 9.** *Let $G = (V, E, w)$ be an edge weighted cocomparability graph, where $w : E \to \mathbb{R}_{>0}$. A maximum weight induced matching on $G$ can be computed in $\mathcal{O}(mn)$ time.*

## 5   Conclusions and perspectives

In this paper, we give a general theorem that shows that a number of vertex ordering characterizations are closed under the operation of taking the square of the line graph. Using the ⋆ and • rules, we get that chordal, threshold, interval, split, and cocomparability graphs all have vertex orderings closed under the $L^2(\cdot)$ operation. This gives in our opinion a natural way to approach this closure under $L^2(\cdot)$; and unifies the results on structural graph classes that have relied on geometric intersection models to show such closure. Furthermore, being able to compute vertex orderings directly can be exploited algorithmically, since algorithms on the graph classes covered often rely on their vertex ordering characterizations. We also show structural results and properties on cocomparability graphs that allow us to compute a maximum weighted induced matching on this graph class in $\mathcal{O}(mn)$ time, an improvement over the best $\mathcal{O}(n^4)$ time algorithm for the unweighted case. A natural question however is whether one can use the vertex orderings $\sigma^*$ of the $L^2(G)$ to compute an induced matching more efficiently for other graph classes, similarly to how we did for cocomparability graphs. We note that the graph classes covered in this work are not necessarily the only ones for which the ⋆, • rules work, thus it's natural to ask what other graph families have this property. In particular, we illustrate our result on graph families with forbidden patterns on three vertices and therefore raise the question of what can be said about forbidden patterns on four or more vertices, but also if other rules exist that preserve orderings in $L^2(G)$.

Another natural question one can raise is whether computing a maximum cardinality induced matching on cocomparability graphs can be done faster than $\mathcal{O}(mn)$ time, especially since computing a maximum cardinality independent set on cocomparability graphs is done with a simple greedy LexDFS based algorithm [11]. LexDFS and LexBFS are graph searching algorithms that have proven powerful on a number of graph families, cocomparability being one of them. We refer the reader to [31, 11, 9, 29, 10, 28] for more on this topic. Unfortunately, one can show that LexDFS cocomparability orderings are not preserved under the ⋆ and • rules, and thus computing such a solution would require computing a LexDFS ordering on $\sigma^*, \sigma^\bullet$. Such an algorithm exists and runs in linear time [22], but it would be linear in the size of $L^2(G)$, thus not in $\mathcal{O}(m+n)$ time. Similarly, LexBFS cocomparability orderings are not preserved under the ⋆ and • rules. We ask the question whether one can come up with a different rule that preserves LexDFS and/or LexBFS cocomparability orderings on $L^2(G)$ without computing the square edges. Such a technique was successfully used with LexBFS on chordal graph in [3].

Lastly, we raise the question of whether $\sigma^*, \sigma^\bullet$ can lead to efficient algorithms to compute a strong edge colouring for these graph classes. Recall that a strong edge colouring is the partitioning of $G$ into induced matchings, and thus the partitioning of $L^2(G)$ into independent sets. The strong chromatic number of $G$ is the size of a minimum strong edge colouring of $G$. It is thus easy to see that the strong chromatic number of $G$ is just $\chi(L^2(G))$. Since the graph families we presented are perfect, their chromatic number can be computed in polynomial time. In fact for many graph families, it is done in linear time, and it often relies on the vertex ordering characterization of the graph class. Since a vertex ordering of $L^2(G)$ can be computed in linear time given $\sigma$, we ask whether $\sigma^*, \sigma^\bullet$ can be used to compute $\chi(L^2(G))$, without computing the edges of $L^2(G)$.

## References

1   Hari Balakrishnan, Christopher L. Barrett, V. S. Anil Kumar, Madhav V. Marathe, and Shripad Thite. The distance-2 matching problem and its relationship to the mac-layer capacity of ad hoc wireless networks. *IEEE Journal on Selected Areas in Communications*, 22(6):1069–1079, 2004.

**2**  Vincenzo Bonifaci, Peter Korteweg, Alberto Marchetti-Spaccamela, and Leen Stougie. Minimizing flow time in the wireless gathering problem. *ACM Trans. Algorithms*, 7(3):33:1–33:20, 2011.

**3**  Andreas Brandstädt and Chính T. Hoàng. Maximum induced matchings for chordal graphs in linear time. *Algorithmica*, 52(4):440–447, 2008.

**4**  Andreas Brandstädt, Van Bang Le, and Jeremy P. Spinrad. *Graph Classes: A Survey.* Society for Industrial and Applied Mathematics, 1999.

**5**  Kathie Cameron. Induced matchings. *Discrete Applied Mathematics*, 24(1-3):97–102, 1989.

**6**  Kathie Cameron. Induced matchings in intersection graphs. *Discrete Mathematics*, 278(1-3):1–9, 2004.

**7**  Kathie Cameron, R. Sritharan, and Yingwen Tang. Finding a maximum induced matching in weakly chordal graphs. *Discrete Mathematics*, 266(1-3):133–142, 2003.

**8**  Jou-Ming Chang. Induced matchings in asteroidal triple-free graphs. *Discrete Applied Mathematics*, 132(1-3):67–78, 2003.

**9**  Pierre Charbit, Michel Habib, Lalla Mouatadid, and Reza Naserasr. Towards A unified view of linear structure on graph classes. *CoRR*, abs/1702.02133, 2017.

**10**  Derek G. Corneil, Barnaby Dalton, and Michel Habib. Ldfs-based certifying algorithm for the minimum path cover problem on cocomparability graphs. *SIAM J. Comput.*, 42(3):792–807, 2013.

**11**  Derek G. Corneil, Jérémie Dusart, Michel Habib, and Ekkehard Köhler. On the power of graph searching for cocomparability graphs. *SIAM J. Discrete Math.*, 30(1):569–591, 2016.

**12**  Peter Damaschke. *Forbidden Ordered Subgraphs.* Topics in Combinatorics and Graph Theory: Essays in Honour of Gerhard Ringel. Physica-Verlag HD, 1990.

**13**  William Duckworth, David Manlove, and Michele Zito. On the approximability of the maximum induced matching problem. *J. Discrete Algorithms*, 3(1):79–91, 2005.

**14**  Shimon Even, Oded Goldreich, Shlomo Moran, and Po Tong. On the np-completeness of certain network testing problems. *Networks*, 14(1):1–24, 1984.

**15**  Martin C. Golumbic. *Algorithmic Graph Theory and Perfect Graphs (Annals of Discrete Mathematics)*, volume 57. North-Holland Publishing Co., 2004.

**16**  Martin Charles Golumbic and Renu C. Laskar. Irredundancy in circular arc graphs. *Discrete Applied Mathematics*, 44(1-3):79–89, 1993.

**17**  Martin Charles Golumbic and Moshe Lewenstein. New results on induced matchings. *Discrete Applied Mathematics*, 101(1-3):157–165, 2000.

**18**  Martin Charles Golumbic, Doron Rotem, and Jorge Urrutia. Comparability graphs and intersection graphs. *Discrete Mathematics*, 43(1):37–46, 1983.

**19**  Pavol Hell, Bojan Mohar, and Arash Rafiey. Ordering without forbidden patterns. In *Algorithms - ESA 2014 - 22th Annual European Symposium, Wroclaw, Poland, September 8-10, 2014. Proceedings*, pages 554–565, 2014.

**20**  Changhee Joo, Gaurav Sharma, Ness B. Shroff, and Ravi R. Mazumdar. On the complexity of scheduling in wireless networks. *EURASIP J. Wireless Comm. and Networking*, 2010, 2010.

**21**  Daniel Kobler and Udi Rotics. Finding maximum induced matchings in subclasses of claw-free and p5-free graphs, and in graphs with matching and induced matching of equal maximum size. *Algorithmica*, 37(4):327–346, 2003.

**22**  Ekkehard Köhler and Lalla Mouatadid. Linear time lexdfs on cocomparability graphs. In *Algorithm Theory - SWAT 2014 - 14th Scandinavian Symposium and Workshops, Copenhagen, Denmark, July 2-4, 2014. Proceedings*, pages 319–330, 2014.

**23** Ekkehard Köhler and Lalla Mouatadid. A linear time algorithm to compute a maximum weighted independent set on cocomparability graphs. *Inf. Process. Lett.*, 116(6):391–395, 2016.

**24** Dieter Kratsch and Lorna Stewart. Domination on cocomparability graphs. *SIAM J. Discrete Math.*, 6(3):400–417, 1993.

**25** Ravi Kumar, Uma Mahadevan, and D. Sivakumar. A graph-theoretic approach to extract storylines from search results. In *Proceedings of the Tenth ACM SIGKDD International Conference on Knowledge Discovery and Data Mining, Seattle, Washington, USA, August 22-25, 2004*, pages 216–225, 2004.

**26** Vadim V. Lozin. On maximum induced matchings in bipartite graphs. *Inf. Process. Lett.*, 81(1):7–11, 2002.

**27** Ross M. McConnell and Jeremy P. Spinrad. Modular decomposition and transitive orientation. *Discrete Mathematics*, 201(1-3):189–241, 1999.

**28** George B. Mertzios and Derek G. Corneil. A simple polynomial algorithm for the longest path problem on cocomparability graphs. *SIAM J. Discrete Math.*, 26(3):940–963, 2012.

**29** George B. Mertzios, André Nichterlein, and Rolf Niedermeier. Linear-time algorithm for maximum-cardinality matching on cocomparability graphs. *CoRR*, abs/1703.05598, 2017.

**30** Hannes Moser and Somnath Sikdar. The parameterized complexity of the induced matching problem in planar graphs. In *Frontiers in Algorithmics, First Annual International Workshop, FAW 2007, Lanzhou, China, August 1-3, 2007, Proceedings*, pages 325–336, 2007.

**31** Donald J. Rose, Robert Endre Tarjan, and George S. Lueker. Algorithmic aspects of vertex elimination on graphs. *SIAM J. Comput.*, 5(2):266–283, 1976.

**32** Larry J. Stockmeyer and Vijay V. Vazirani. Np-completeness of some generalizations of the maximum matching problem. *Inf. Process. Lett.*, 15(1):14–19, 1982.